\documentclass{cimento}
\usepackage{float}
\usepackage{amsmath,amssymb}
\usepackage{graphicx}
\usepackage{mathtools,slashed}
\usepackage{bm}
\usepackage{comment}
\usepackage{xcolor}
\usepackage{subcaption}
\usepackage{array}
\usepackage{multirow}
\usepackage{url}
\usepackage{color}
\usepackage[T1]{fontenc}
\usepackage[latin9]{inputenc}
\usepackage{graphicx}
\usepackage{esint}
\usepackage{atbegshi}
\usepackage{lipsum}

\title{Spectral reconstruction of Euclidean correlator moments in lattice QCD}
\author{John~Bulava \\ for the Baryon Scattering (BaSc) Collaboration}
\instlist{\inst Fakult\"{a}t f\"{u}r Physik und Astronomie, Institut f\"{u}r Theoretische Physik II, Ruhr-Universit\"{a}t Bochum, 44780 Bochum, Germany  }

\begin{document}

\maketitle

\begin{abstract}
	A novel application of lattice QCD spectral reconstruction is presented,
	in which euclidean correlation function data in a fixed time range 
	are used to infer values outside the range, enabling a model-independent 
	investigation of the asymptotic large-time behavior.
	Moments of the correlator are also determined,  
  and reconstructed correlation matrices between different moments are included in a variational optimization similar to the standard Generalized Eigenvalue Problem (GEVP). These ideas are illustrated using a single-nucleon correlation function determined on an $N_{\rm}=2+1$ ensemble of gauge configurations at $m_{\pi} = 200{\rm MeV}$. 
\end{abstract}

In a finite spatial volume, the spectrum of QCD is discrete. The energy gaps suggest a strategy for the study of low-lying states in lattice QCD: take the large-time limit of Euclidean correlation functions to 
suppress unwanted excited state contamination. Solutions of a Generalized Eigenvalue Problem~\cite{Michael:1982gb} can be used to construct correlators asymptotically 
dominated by a single state. 
However, (nearly) all correlation functions suffer from an exponential degradation of the signal-to-noise ratio with increasing Euclidean time~\cite{Parisi:1983ae,Lepage:1989hd}. Unfortunately, in practice the most precise data at early times 
is discarded to fit the large-time region to a few-state ansatz. 
Choosing the lower bound for such fits is often delicate: the statistical error can be decreased at the expense of increased systematic error due to excited state contamination. 
It is possible that this interplay is (at least partly) responsible for the long-standing difficulty in 
reproducing the experimental value for the nucleon axial charge~\cite{Gupta:2021ahb,Barca:2022uhi}. 

A model-independent 
alternative to discarding the early-time data is presented here. It is based on the spectral reconstruction approach proposed by Backus and Gilbert~\cite{BG1} and first 
used in lattice QCD by Hansen, Lupo, and Tantalo~\cite{Hansen:2019idp}.  Ref.~\cite{Bailas:2020qmv} is similar in spirit to this work and is demonstrably of comparable effectiveness~\cite{Barone:2022gkn}. Other approaches to spectral reconstruction in lattice QCD are reviewed in Ref.~\cite{Rothkopf:2022fyo}. The general problem of spectral reconstruction employs 
Euclidean two-point correlator data 
\begin{align}\label{e:c1}
	C(t) = \int {\rm d}\omega \, \rho(\omega) \, {\rm e}^{-\omega t}, \qquad 
	\rho(\omega) = \sum_n A_n \, \delta(\omega - E_n) 
\end{align}
known for integer values of $t/a$ in the range 
$t \in [t_{\rm min}, t_{\rm max} ]$ with statistical errors. The direct 
determination of $\rho(\omega)$ from the correlator data is an ill-posed 
problem, which 
can be ameliorated by instead seeking smeared spectral densities of the form 
	$\tilde{\rho}[f] = \int {\rm d}\omega f(\omega)\, \rho(\omega)$ with a particular smearing kernel $f(\omega)$ specified 
\emph{a priori}.  
Previous applications of Ref.~\cite{Hansen:2019idp} use $f(\omega)$ to 
approximate the Dirac-$\delta$ distribution~\cite{Bulava:2021fre},  
Heaviside step function~\cite{Gambino:2022dvu}, or implement the principle-value prescription~\cite{Frezzotti:2023nun}. This work however determines correlator moments, defined as 
\begin{align}\label{e:corr}
	D(\alpha,\tau) = \int {\rm d}\omega \, \rho(\omega) \, \omega^{\alpha} \, {\rm e}^{-\omega \tau}. 	
\end{align}
 The different symbol emphasizes that $\tau$ need not be
constrained to the values of $t$ provided by the data, suggesting that the 
temporal resolution may be increased with non-integer $\tau/a$ and 
the asymptotic limit probed with $\tau > t_{\rm max}$. Furthermore, taking the (rational) power $\alpha$ different from zero suppresses or enhances excited 
state contamination. Correlators with different $\alpha$ may be viewed as employing different interpolating operators, which can accordingly form a correlation matrix for use in a Generalized eigenvalue problem (GEVP). 

The reconstruction approach employed here is detailed in Ref.~\cite{Hansen:2019idp}, with further discussion of the reliable estimation 
of statistical and systematic errors in 
Refs.~\cite{Bulava:2021fre,Bulava:2023mjc}. Consider an estimator for $D(\alpha,\tau)$ from Eq.~\ref{e:corr} which is a linear combination of all input correlator timeslices
\begin{gather}
	\hat{D}(\alpha,\tau) = \sum_{t = t_{\rm min}}^{t_{\rm max}} 
	g_t(\alpha,\tau)\, C(t) 
	 \equiv \int{\rm d}\omega \, \hat{f}(\omega) \, 
	\rho(\omega).
\end{gather}
Evidently $\hat{D}(\alpha,\tau)$ is a smeared 
spectral density smeared with the kernel $\hat{f}(\omega) =   \sum_t g_t(\alpha,\tau)\,{\rm e}^{-\omega t}$, in contrast to $D(\alpha,\tau)$ which is smeared with $f(\omega) = \omega^\alpha \, {\rm e}^{-\omega \tau}$.  Any difference between $\hat{f}(\omega)$ and 
$f(\omega)$ is a systematic error in the estimator $\hat{D}(\alpha,\tau)$. The 
systematic and statistical errors are quantified by 
\begin{gather}\label{e:afunc}
	A[g] = \int_{\omega_0}^{\infty} {\rm d}\omega \, \left\{f(\omega) - \hat{f}(\omega) \right\}^2, \qquad B[g] = \sum_{tt'} g_tg_{t'} \, {\rm Cov}\left\{ C(t), C(t') \right\}, 
\end{gather}
respectively. 
The coefficients  $\{g_t\}$ are chosen to balance these two considerations by  
minimizing $G_{\lambda}[g] = (1-\lambda)A[g] + \lambda B[g]$ 
for a particular $\lambda \in [0,1]$. 
Small $\lambda$
results in a reconstructed kernel $\hat{f}(\omega)$ close to the 
desired one $f(\omega)$, but at the expense of 
large statistical error. Large $\lambda$ gives a statistically precise result, 
but with a large systematic error due to the difference between 
$\hat{f}(\omega)$ and $f(\omega)$. As is customary in lattice QCD data 
analysis, an ideal $\lambda$ is sought in the statistics-limited regime 
wherein the systematic error is 
reliably smaller than the statistical error.

\begin{figure}
	\includegraphics[width=0.49\textwidth]{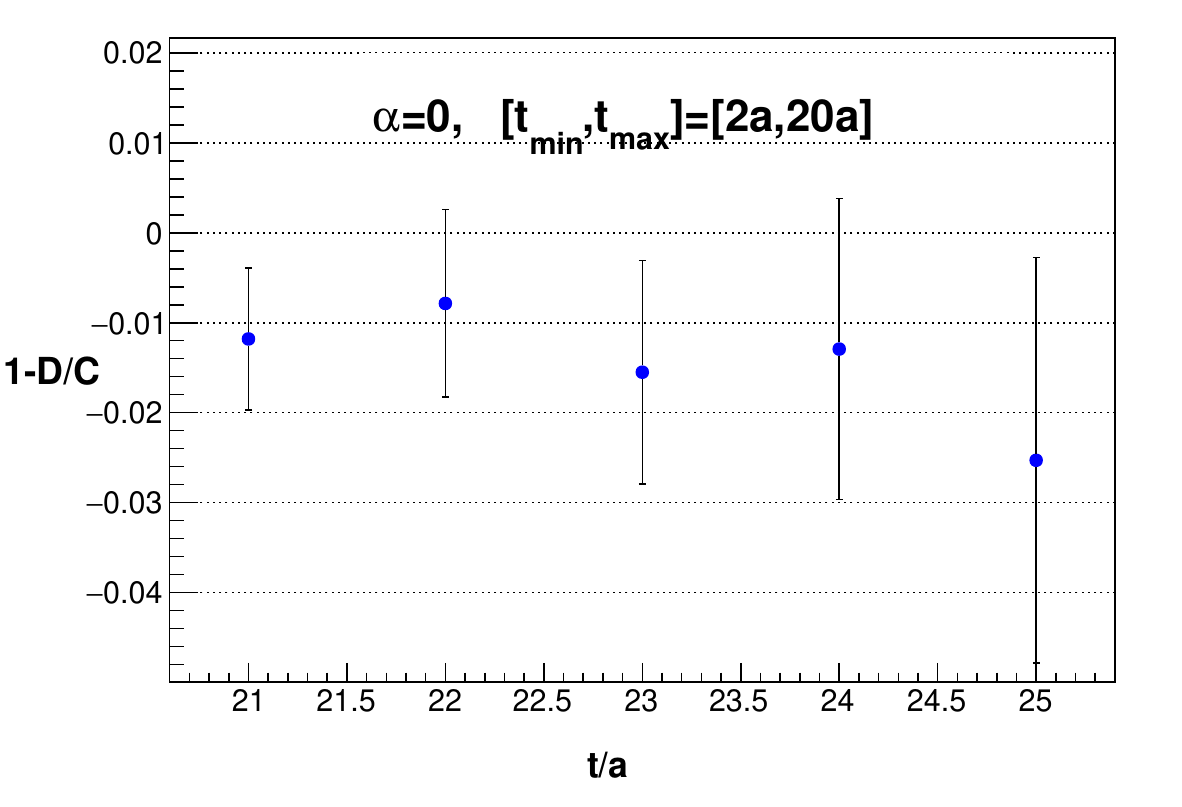}
	\includegraphics[width=0.49\textwidth]{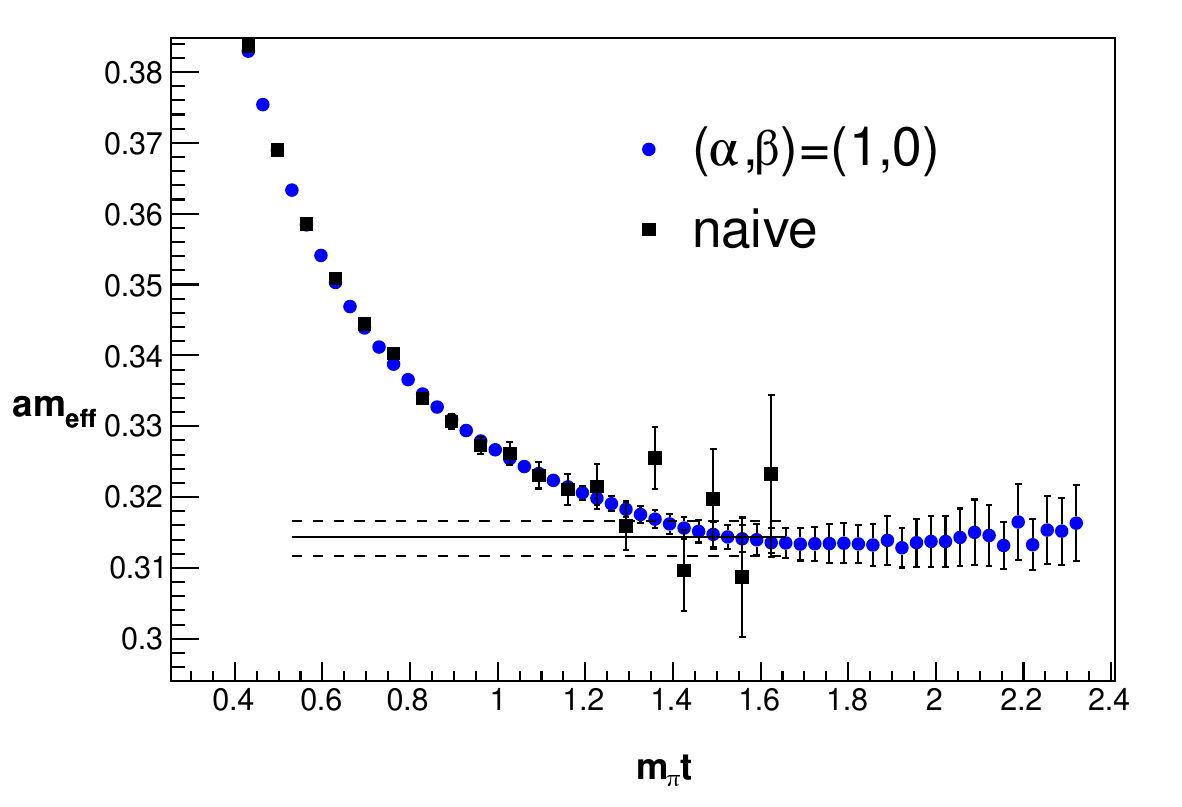}
	\caption{\label{f:test}Two tests of the correlator reconstruction procedure. 
	{\bf Left}: using correlator data from times $t/a = [2,20]$ to infer $D(0,t)$ at later times and compare with direct measurements of $C(t)$. The error on the relative difference between these two estimators is computed using the bootstrap procedure. {\bf Right}: using times $t/a = [2,25]$ to infer the effective mass using Eq.~\ref{e:meff} with $(\alpha,\beta)=(1,0)$. The naive forward-difference definition is also shown for comparison. The horizontal band is the result of a two-state fit (from Ref.~\cite{Bulava:2022vpq}) over the horizontal 
	range shown.}
\end{figure}

To test these ideas, a single-nucleon correlator from an $N_{\rm f}=2+1$ ensemble of gauge configurations with $m_{\pi}=200\,{\rm MeV}$ is employed. This correlator is computed over the time range  $[t_{\rm min}, t_{\rm max}] = [2a, 25a]$ and was used to determine the nucleon mass $m_{\rm N}$ in Ref.~\cite{Bulava:2022vpq}. Further details of the interpolating operators and measurement procedure can be found there. For the reconstruction, the lower bound of the integration in Eq.~\ref{e:afunc} is fixed at $a\omega_0 = 0.3$, and the optimal $\lambda$ chosen as in Ref.~\cite{Bulava:2023mjc} by demanding that the variation of the estimator among a set of reconstructions which impose different constraints is three times smaller than the statistical error. As a first test of the procedure, 
a reconstruction is preformed using timeslices $[t_{\rm min}, t_{\rm max}] = [2a,20a]$ to infer $t=21a,\dots,25a$. The results of this test are shown in the left panel of Fig.~\ref{f:test}, which demonstrates that the standard estimator for the correlator and $D(0,\tau)$ agree within $1.5\sigma$.  

Consider now the family of effective masses   
\begin{gather}\label{e:meff}
	m_{\rm eff}(\alpha, \beta|\tau) = \left\{ \frac{D(\alpha, \tau)}{D(\beta,\tau)}\right \}^{1/(\alpha-\beta)} 
\end{gather}
for which $\lim_{\tau \rightarrow \infty}  m_{\rm eff}(\alpha,\beta|\tau) = m_{\rm N}\times \left\{1 + {\rm O}({\rm e}^{-2m_{\pi}\tau})\right\}$.  Due to the finite spatial volume, the lowest non-interacting $p$-wave state has an energy close to $m_{\rm N}+2m_{\pi}$. The case $(\alpha,\beta)= (1,0)$ coincides with the standard forward-difference effective mass up to $O(a^2)$ 
while different $(\alpha,\beta)$ have varying excited state contamination. A second test of the reconstruction procedure is shown in the right panel of Fig.~\ref{f:test}, which compares the forward difference effective mass with $(\alpha,\beta)=(1,0)$. The reconstructed effective mass is considerably more precise at 
later times, and asymptotically agrees with the value obtained from a two-state fit. It should be emphasised, however, that the spectral reconstruction approach does not impose a model for the time dependence.

\begin{figure}
	\centering
	\includegraphics[width=0.6\textwidth]{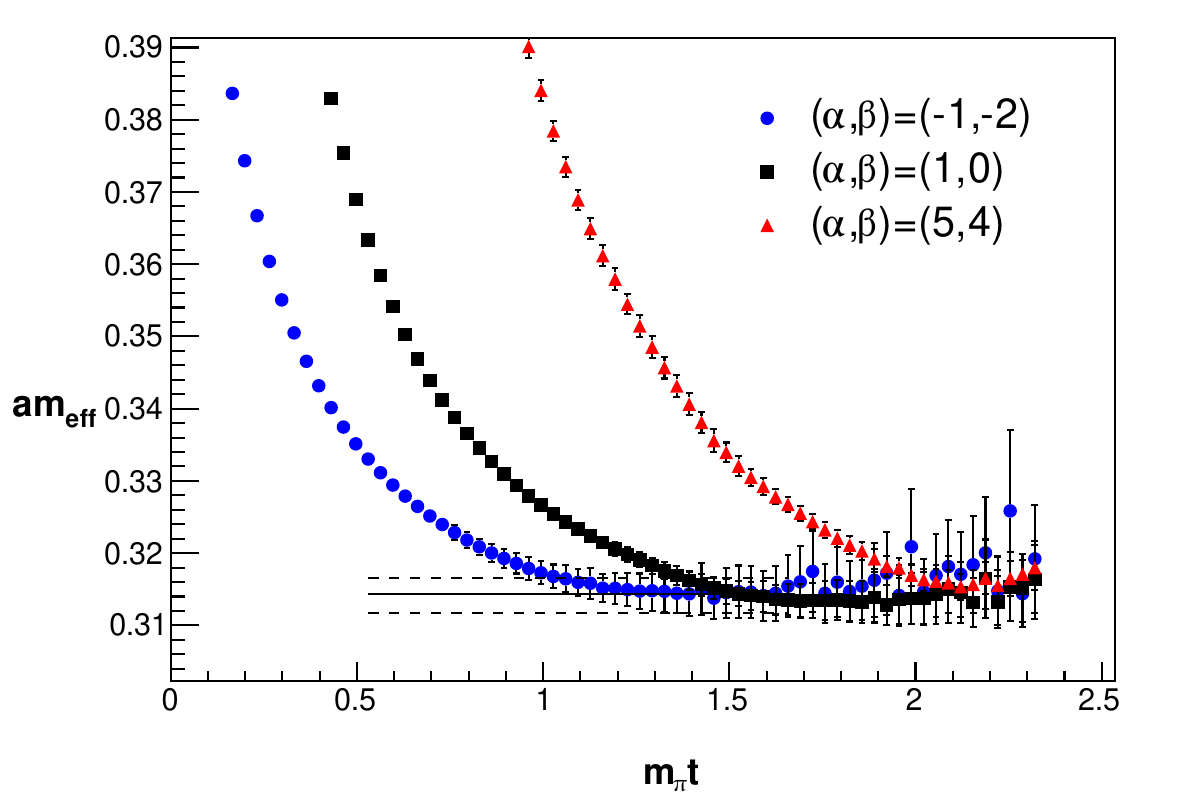}
	\caption{\label{f:emass} Different definitions of the effective mass by varying $\alpha$ and $\beta$ in Eq.~\ref{e:meff}. In this manner the excited state contamination can be suppressed or enhanced relative to the conventional definition, which employs $(\alpha, \beta)=(1,0)$. }
\end{figure}
\begin{figure}
	\includegraphics[width=0.49\textwidth]{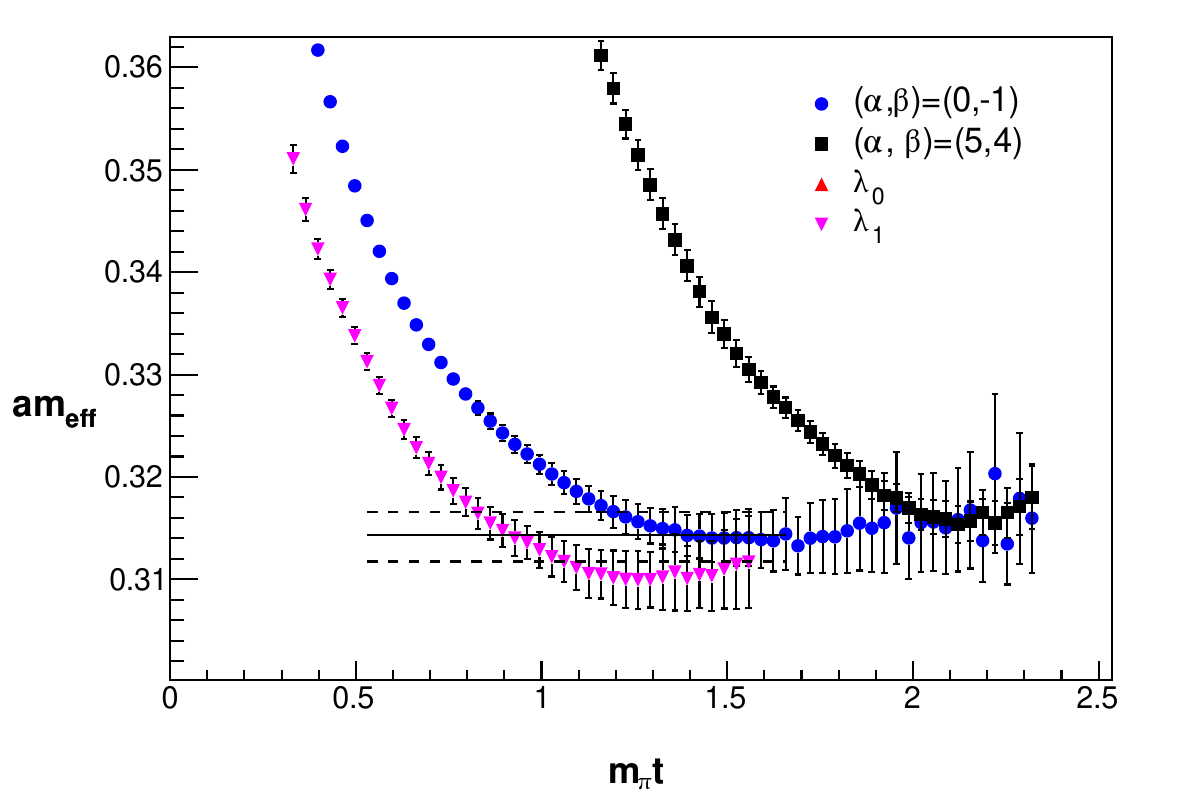}
	\includegraphics[width=0.49\textwidth]{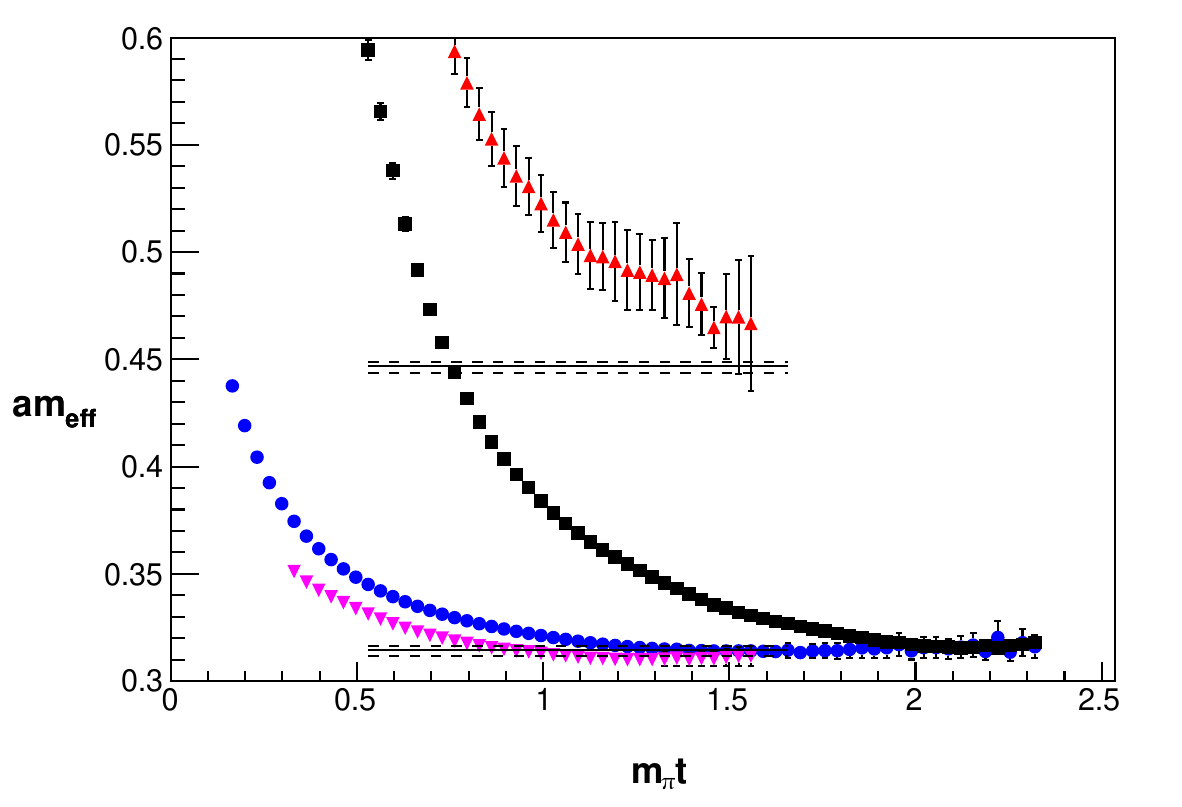}
	\caption{\label{f:gevp} The eigenvalue $\lambda_n(t)$ from a two-by-two GEVP as in Eq.~\ref{e:gevp} with $\{\alpha_1,\alpha_2\}=\{-0.5,2\}$. The lower horizontal band is the two-state fit from Ref.~\cite{Bulava:2022vpq}, as in 
	Fig.~\ref{f:test}. The right panel is zoomed out to show $\lambda_1(t)$, and the upper horizontal band is $m_{\rm N}+2m_{\pi}$.}
\end{figure}

Fig.~\ref{f:emass} shows various $(\alpha,\beta)$. A significant 
enhancement or reduction of the excited state contamination is evident similar to a variation in the level of quark field smearing, but here achieved 
for a single correlator. This naturally suggests employing different moments 
in a correlation matrix $A_{ij}(t) = D(\alpha_i + \alpha_j, t)$. Rather than 
the standard GEVP, which employs a second metric timeslice $t_0$, the variation optimization is applied directly to $B_{ij}(t) = D(\alpha_i+\alpha_j+1, t)$, 
using $A(t)$ as a metric. This `equal-time' GEVP is therefore  
\begin{gather}\label{e:gevp}
	B(t)v_n(t) = \lambda_n(t) A(t) v_n(t), 
\end{gather}
where the eigenvalues $\lambda_n(t)$ approach directly the states of interest for large $t$. The results 
from a two-dimensional equal-time GEVP are shown in Fig.~\ref{f:gevp}, which 
show a further reduction in excited state contamination compared to the input 
`diagonal' effective masses. Furthermore, a rough estimate of the first excited state is provided with an energy near $m_{\rm N}+2m_{\pi}$. Interactions likely shift the multi-hadron excited states significantly from these naive expectations, however. 

In conclusion, the reconstruction of correlator moments presented here provides model-independent determinations of the correlator at arbitrary 
Euclidean time separations. Unlike few-state fits, these estimates employ 
the entire range of correlator data, including those at precise early times.
However, no miracle has been achieved: the asymptotic behaviour shown in Figs.~\ref{f:test},~\ref{f:emass}, and~\ref{f:gevp} is consistent with a two-state fit and has a similar statistical precision. Nonetheless, it serves as a valuable model-independent 
confirmation of the two-state ansatz over the limited time range of input correlator data.  
In addition to the investigation of large-time asymptotics detailed here, the model-independent interpolation of correlator data may be useful in the comparison of lattice QCD vector-vector correlators and smeared experimental $R$-ratio data.  
It is likely that this approach can also be extended to treat the simultaneous reconstruction of multiple correlation functions in analogy to the ratios constructed to determine energy differences, as well as three-point correlation functions.  

\acknowledgments
I thank my co-authors of Ref.~\cite{Bulava:2022vpq} and my colleagues in the 
Baryon Scattering Collaboration (BaSc) for sharing the single-nucleon correlator used here.

\bibliography{latticen}

%


\end{document}